\documentclass[letterpaper,pre,twocolumn,showpacs]{revtex4}

\usepackage{graphicx}

\newcommand{\eg} {{e.g., }}
\newcommand{\etam} {\eta_{\rm M}}
\newcommand{\etat} {\eta_{\rm T}}
\newcommand{\etav} {\eta_{\rm V}}
\newcommand{\etatv} {\eta_{\rm T,V}}
\newcommand{\half} {\frac{1}{2}}
\newcommand{\gammael} {\gamma_{\rm el}}
\newcommand{\ie} {{i.e., }}

\begin{document}

\title
{Nanoscale surface relaxation of a membrane stack}

\author{Hamutal Bary-Soroker}
\altaffiliation[Present address: ]{Department of Condensed Matter Physics,
Weizmann Institute of Science, Rehovot 76100, Israel.}
\affiliation{School of Physics \& Astronomy, Raymond \& Beverly
  Sackler Faculty of Exact Sciences, Tel Aviv University, Tel Aviv
  69978, Israel}
\author{Haim Diamant}
\email{hdiamant@tau.ac.il} 
\affiliation{School of Chemistry, Raymond \& Beverly Sackler Faculty
  of Exact Sciences, Tel Aviv University, Tel Aviv 69978, Israel}

\date{May 29, 2007}

\begin{abstract}
  Recent measurements of the short-wavelength ($\sim 1$--$100$ nm)
  fluctuations in stacks of lipid membranes have revealed two distinct
  relaxations: a fast one (decay rate of $\sim 0.1$ ns$^{-1}$), which
  fits the known baroclinic mode of bulk lamellar phases, and a slower
  one ($\sim 1$--$10$ $\mu$s$^{-1}$) of unknown origin. We show that
  the latter is accounted for by an overdamped capillary mode,
  depending on the surface tension of the stack and its anisotropic
  viscosity. We thereby demonstrate how the dynamic surface tension of
  membrane stacks could be extracted from such measurements.
\end{abstract}


\maketitle

Self-assembled stacks of membranes are encountered in various
industrial and biological systems. They consist of parallel bilayers
of amphiphilic molecules separated by microscopic layers of solvent
--- a structure with the symmetry of a smectic A liquid crystal
\cite{LC}. Such stacks form lyotropic lamellar phases \cite{lamellar},
on which many cleaning and cosmetic products are based. Lamellar
bodies are found also in the lung \cite{lung} and as multilayer
vesicles (``onions'') \cite{onion}. Membrane stacks made of
phospholipids have been widely used to study properties of
biological membranes, whereby the large number of identical, equally
spaced membranes helps enhance the signal and allows the study of
membrane--membrane interactions (\eg \cite{Parsegian}).

The elasticity of membrane stacks is equivalent to that of
single-component (thermotropic) smectics \cite{LC} and has been
extensively studied. The elastic moduli of the stack can be extracted
from its equilibrium fluctuations using, \eg x-ray line shape analysis
\cite{Safinya}. By contrast, the hydrodynamics of membrane stacks 
\cite{Brochard,Nallet}, because of their two micro-phase-separated 
components, differs from that of thermotropic smectics \cite{Martin}.
An additional hydrodynamic mode appears --- the baroclinic (slip) mode
--- along with a unique dissipation mechanism, in which the membranes
and solvent layers develop different average velocities
\cite{Brochard}. Experimental studies of hydrodynamic modes in
membrane stacks have been rather scarce, the prevalent technique being
dynamic light scattering \cite{Nallet}, whose spatial resolution is
limited by the wavelength of light.

In a recent experiment using neutron spin-echo spectrometry,
Rheinst\"adter, H\"au{\ss}ler, and Salditt (RHS) have provided a first
look at the relaxation of membrane stacks at short wavelength
($1$--$100$ nm) and short time ($1$--$10^3$ ns) \cite{RHS}. Their
system consisted of several thousands of
dimyristoylphosphatidylcholine (DMPC) phospholipid bilayers,
self-assembled into a stack of $d\sim 5$ nm periodicity.  The system
was studied at temperatures above and below the lipid melting point,
corresponding to fluid and gel-like membranes, respectively. In both
cases the measured dynamics consisted of two distinct exponential
relaxations. The dispersion relation of the faster relaxation (decay
rate of $\sim 0.1$ ns$^{-1}$) could be well fitted in the
fluid-membrane case to that of the baroclinic mode of a bulk lamellar
phase \cite{Ribotta}, while the slower mode (decay rate of $\sim
1$--$10$ $\mu$s$^{-1}$) was left unexplained. We demonstrate below
that this slower relaxation is well accounted for by a surface mode,
\ie a perturbation which is localized within a finite penetration
depth from the surface of the stack.

In a recent publication \cite{epl} we have addressed the surface
dynamics of membrane stacks, highlighting the qualitative differences
from the surface dynamics of both simple liquids and thermotropic
smectics \cite{Romanov}. These differences arise from the slip
dissipation mechanism, which is absent in simple liquids and
thermotropic smectics but is usually dominant in lyotropic lamellar
phases. Although the formulation in Ref.\ \cite{epl} is general, its
analysis is focused on a very different domain (larger wavelengths and
slower rates) from that sampled by RHS.  In that domain the slip
dissipation dominates and, consequently, the surface relaxation is
governed by an overdamped diffusive mode, whose decay rate $\Gamma$
increases quadratically with the wavevector $q$. In this Brief Report
we present a slight adaptation of that theory for a large-$q$,
high-$\Gamma$ regime such as that of RHS.

The general surface dynamics of membrane stacks is quite complex,
depending on several restoring and dissipation mechanisms
\cite{epl}. Three moduli are associated with the restoring forces: the
compression modulus $B$, bending modulus $K$, and surface tension
$\gamma$. Viscous dissipation is characterized (in the limit of
incompressible flow) by three viscosity coefficients \cite{Brochard},
denoted $\etam$, $\etat$, and $\etav$. The coefficient $\etam$,
associated with differences in the lateral velocity across layers
(sliding viscosity), is much smaller than the other two, which
correspond to the viscous response to deformations of the lipid
membranes. We use the parameter $\Theta=2(\etat+\etav)/\etam$ to
characterize this viscosity anisotropy; it is typically of order
$10^2$--$10^3$
\cite{Brochard,Colin}. The aforementioned slip motion requires another
transport coefficient \cite{Brochard}, $\mu\simeq d^2/(12\eta_0)$,
where $\eta_0$ is the viscosity of the solvent (water) layer.

\begin{figure}[tbh]
\centerline{\resizebox{0.47\textwidth}{!}
{\includegraphics{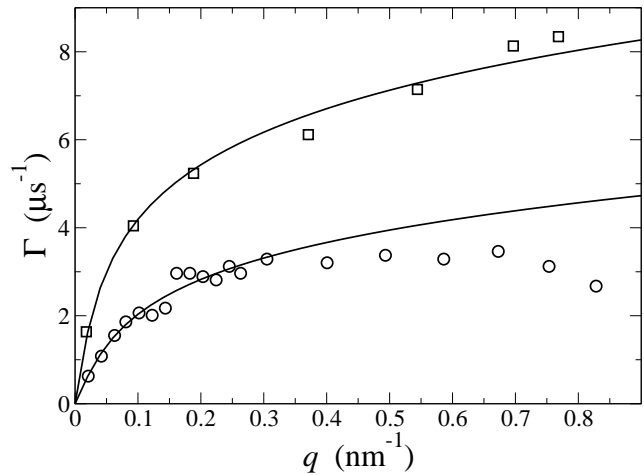}}}
\caption[]{Dispersion relations for the slower relaxation mode of stacks
of DMPC lipid membranes at 30$^\circ$C (circles) and 19$^\circ$C
(squares). (Data taken from Ref.\ \cite{RHS}.) The solid lines are
fits to Eq.\ (\ref{dispersion}) with $d=5.4$ nm, $\etam=0.016$ Pa s,
$\Theta=110$, and $\gamma=5.4$ mN/m (lower curve); $d=5.6$ nm,
$\etam=0.016$ Pa s, $\Theta=350$, and $\gamma=28$ mN/m (upper curve).
The values of $d$ and $\etam$ are taken from Ref.\ \cite{RHS};
$\Theta$ and $\gamma$ are fitting parameters.}
\label{fig_1}
\end{figure}

In view of this richness it is helpful to begin by identifying the
dominant contributions to the slower mode of Ref.\ \cite{RHS}. First,
for the typical parameters of that case --- $q\sim 10^{-1}$ nm$^{-1}$,
$\Gamma\sim 1$ $\mu$s$^{-1}$, $\etam\sim 10^{-2}$ Pa~s, and mass
density $\rho\sim 1$ g/cm$^3$ --- one gets a negligible Reynolds
number, ${\rm Re}\sim\rho\Gamma/(\etam q^2)\sim 10^{-5}$, implying
that inertial modes \cite{Romanov} are irrelevant in the current
case. Second, to determine the dominant dissipation mechanism one
should compare the friction due to slip, $\mu^{-1}v$ ($v$ being a
characteristic relative velocity), with that due to viscous stresses,
$\etatv q^2 v$, \ie the dimensionless parameter $S=(\etam\mu
q^2)^{-1}$ is to be compared with $\Theta$ \cite{epl}. We find $S\sim
10\ll\Theta$. Thus, unlike the mode focused on in Ref.\ \cite{epl}, in
the current large-$q$ case viscous dissipation is dominant. Finally,
the relative importance of the three restoring mechanisms depends not
only on the surface perturbation wavevector $q$ but also on its
penetration depth $\alpha^{-1}$. Since the value of $\alpha$ is
unknown {a priori}, all three mechanisms should be considered in
principle. However, to keep the analysis as simple as possible we
shall assume that the surface tension is the dominant factor.  This
ansatz is motivated by the experimental fact that the rate of RHS's
slower mode is linear in $|q|$ at small $q$ (see Fig.\ \ref{fig_1});
the way to get such a linear overdamped dispersion relation is to
balance a surface tension stress against a viscous one, $\gamma q^2
u\sim\eta q\Gamma u$ ($u$ being the amplitude of the surface
deformation). We will return to the consistency of this assumption
later on.

The continuum theory formulated in Ref.\ \cite{epl} is valid for
wavelengths much larger than the inter-membrane spacing, $qd\ll
1$. RHS's experiment, however, samples the range $0.1<qd<4$.  To
obtain an extrapolation of the analysis to large $q$ we introduce one
last modification to the theory --- the distance $z$ from the surface
into the stack is discretized, $z\rightarrow -dn$ $(n=0,1,2,\ldots)$,
turning the differential equations of Ref.\ \cite{epl} into
finite-difference ones (similar to the analysis of high-$q$ acoustic
modes in a crystal). The lateral position $x$ parallel to the
membranes is kept continuous, and we consider, for simplicity, a
surface perturbation which is uniform in the second lateral direction
$y$.
 
Within these assumptions Eq.\ (11) of Ref.\ \cite{epl} yields the
following surface mode for the vertical displacements of the membranes,
$u_n(x,t)$:
\begin{eqnarray}
  u_n &=& (C_+ e^{-\alpha_+dn} + C_- e^{-\alpha_-dn})e^{iqx-\Gamma t}
 \nonumber\\
  \alpha_\pm &=& \frac{2}{d}\sinh^{-1}\left(\half\Theta^{\pm 1/2}|q|d \right).
\label{mode}
\end{eqnarray}
For sufficiently small $q$ ($qd\ll\Theta^{-1/2}$) the spatial decay
coefficients are $\alpha_\pm\simeq\Theta^{\pm 1/2}|q|$, \ie the mode
contains two terms of disparate penetration depths,
$\alpha_-^{-1}\gg\alpha_+^{-1}$. (A qualitatively similar result was
obtained for the surface mode analyzed in Ref.\ \cite{epl}, yet in
the current case the origin of the two differing penetration depths is
the large viscosity anisotropy rather than the strong slip
dissipation.) In the other limit of $qd\gg\Theta^{1/2}$, as expected,
both contributions become localized within a distance of order $d$
from the surface, $\alpha_\pm\simeq (2/d)\ln(\Theta^{\pm 1/2}|q|d)$.

The dispersion relation $\Gamma(q)$ is set by the boundary conditions
for the stress tensor at the stack surface, as summarized in Eq.\ (13)
of Ref.\ \cite{epl}. Substituting in that equation the expressions for
$\alpha_\pm$ obtained above, we get, within the same approximations,
\begin{eqnarray}
  &&\Gamma(q) = \frac{2\gamma}{\Theta\etam d}\times\nonumber\\
  &&\left[\sinh^{-1} \left(\half\Theta^{1/2}|q|d \right) + 
  \sinh^{-1} \left(\half\Theta^{-1/2}|q|d \right) \right].
\label{dispersion}
\end{eqnarray}
Equation (\ref{dispersion}) is the main result of our current
analysis. For large wavelengths this dispersion relation becomes
\begin{equation}
  \Gamma (qd\ll\Theta^{-1/2}) \simeq 
  \frac{\gamma}{2[\etam(\etat+\etav)/2]^{1/2}}|q|.
\label{dispersion1}
\end{equation}
Equation (\ref{dispersion1}) is equivalent to the dispersion relation
of an overdamped capillary mode at the surface of a simple liquid having
effective viscosity $\eta_{\rm eff}=[\etam(\etat+\etav)/2]^{1/2}$.  In
the opposite, short-wavelength limit we get
\begin{equation}
  \Gamma(qd\gg\Theta^{1/2}) \simeq \frac{\gamma}{(\etat+\etav)d/2}\ln(|q|d).
\end{equation}
In this quasi-two-dimensional limit the dependence on the smaller
(sliding) viscosity, $\etam$, disappears, and an effective
two-dimensional viscosity emerges, $\eta_{\rm 2D}=(\etat+\etav)d/2$
\cite{ft_ZG}.

Figure \ref{fig_1} shows fits of the dispersion relations for the
slower mode, as measured by RHS, to Eq.\ (\ref{dispersion}) 
\cite{ft_notfitted}. (The measurements for $q>0.5$ nm$^{-1}$ 
are considered less reliable due to scattering by defects in the stack
\cite{RHS}.) The stack periodicity was measured as $d=5.4$ and $5.6$
nm at temperatures $T=30^\circ$C (fluid membranes) and $19^\circ$C
(gel-like membranes), respectively \cite{RHS}. The value of the
sliding viscosity at $30^\circ$C, $\etam=0.016$ Pa~s, was
independently found from a fit of the faster mode \cite{RHS}. We are
thus left with two fitting parameters in Eq.\ (\ref{dispersion}),
$\Theta$ and $\gamma$. For the fluid-membrane case we find
$\Theta=110$ and $\gamma=5.4$ mN/m. It should be stressed that having
two free parameters does not allow for accurate determination of both,
and these values should be regarded merely as rough
estimates. Nonetheless, the fitted values are of the correct
scale. The value for $\Theta$ implies $\etatv\sim\Theta\etam\sim 1$
Pa~s, \ie a viscosity 3 orders of magnitude larger than that of water,
which matches the typical effective viscosity of lipid membranes
\cite{Sens}. It also implies an effective two-dimensional viscosity
$\eta_{\rm 2D}\sim\etatv d\sim 10^{-9}$--$10^{-8}$ Pa~s~m, which
agrees well with measurements of the surface viscosity of fluid DMPC
membranes \cite{Dimova}.

The applicability of the theory to stacks of solid, gel-like membranes
should be questioned, as such stacks have additional intra-membrane
elasticity. The same concern, in fact, should be raised regarding the
fluid-membrane case as well, since at the high frequencies considered
here the individual membranes are expected to have a viscoelastic
response. The fits obtained in Fig.\ \ref{fig_1} (in particular, the
linear behavior for small $q$) suggest, however, that these additional
restoring forces are negligible compared to the surface tension and do
not affect the surface relaxation. The fit for $T=19^\circ$C yields
significantly larger values for both the viscosity anisotropy and the
surface tension, $\Theta=350$ and $\gamma=28$ mN/m, which is the
expected trend for stiffer membranes \cite{ft_diverge}. (In the fit we
have assumed that the sliding viscosity $\etam$ does not change much
with temperature.)

The elasticity of membrane stacks gives rise to an effective static
surface tension, $\gammael=(KB)^{1/2}$ \cite{LC,Durand,Fournier}. The
values of $K$ and $B$ in the fluid-membrane state were extracted by
RHS from the fit of the faster mode as $K\simeq 1.15\times 10^{-11}$ N
(corresponding to a membrane bending modulus $\kappa=Kd\simeq 14.8$
$k_{\rm B}T$) and $B\simeq 1.08\times 10^7$ Pa. This yields
$\gammael\simeq 11.1$ mN/m, which is of the same order of magnitude as
the high-frequency surface tension fitted above \cite{ft_static}.

We now return to check the self-consistency of our assumptions.
First, for a mode to be a surface one its penetration depth must be
smaller than the total thickness of the sample. The penetration depth
found in Eq.\ (\ref{mode}) is $\alpha_-^{-1}<\Theta^{1/2}/q\lesssim
10^2$ nm, which is at least 1--2 orders of magnitude smaller than the
thickness of RHS's films ($\sim 10$ $\mu$m). Second, for surface
tension to be the dominant restoring force, one should have
$\gamma>\gammael$. This condition can be obtained rigorously
\cite{Hamutal} but is also realized upon demanding that the stress
arising from surface tension, $\gamma q^2\alpha u$, be larger than
both the compression one, $B\alpha^2 u$, and the bending one, $Kq^4
u$. As described above, we actually have $\gamma\sim\gammael$ and,
thus, the assumption can be only marginally fulfilled. Moreover, the
omission of the bending terms requires also that $Kq^2/\etam$ be
smaller than $\Gamma$ \cite{epl}, which is satisfied only for the
lowest end of the sampled $q$ range, $q\lesssim 0.1$ nm$^{-1}$. The
apparent success of the simplified theory over the extended $q$ range
(Fig.\ \ref{fig_1}), therefore, is somewhat surprising. We note that
the stacks of RHS are densely packed. The thickness of a DMPC bilayer
at $30^\circ$C is $4.5$ nm \cite{Kucerka}, implying that the solvent
layers in-between membranes are only $1$ nm thick. For such density
and high-$q$ surface perturbations the stack might not follow the
usual description of linear smectic elasticity but respond merely as
an anisotropic viscous liquid with surface tension.

In summary, the relaxation of nanoscale fluctuations in finite
membrane stacks seems to occur via two distinct overdamped modes --- a
bulk baroclinic mode and a slower surface mode. The dispersion
relation of the surface mode provides access to the dynamic surface
tension of the stack, which should be hard to measure
otherwise. Supplementing such an experiment with measurements at
larger wavelengths (\eg using dynamic light scattering), yielding a
value for $\Theta$, may allow the accurate extraction of the
dynamic surface tension.

\begin{acknowledgments}
  We are indebted to Maikel Rheinst\"adter and Tim Salditt for sharing
  their experimental results with us prior to publication. This work
  was supported in part by the US--Israel Binational Science
  Foundation (2002271).
\end{acknowledgments}




\begin{thebibliography}{99}
\bibitem{ft_notfitted}
These measurements were not compared with a theory before.  RHS used
the hydrodynamic theory of bulk lamellar phases to fit the data of the
faster mode alone \cite{RHS}.
\bibitem{ft_diverge}
While the effective viscosity of individual membranes should diverge
in the gel-like state \cite{Dimova}, $\etatv$ should not, as they
characterize the three-dimensional stack, which is still overall
liquid \cite{Brochard}.
\end{thebibliography}
\end{document}